\documentclass[10pt,conference,letterpaper]{IEEEtran}

\usepackage{amsmath}
\usepackage{amsfonts}
\usepackage[linesnumbered,ruled,vlined]{algorithm2e}
\usepackage{graphicx}
\usepackage{multirow}
\usepackage{subfig}

\usepackage[colorlinks,linkcolor=RoyalBlue,citecolor=webgreen]{hyperref}
\usepackage{siunitx}
\usepackage[dvipsnames]{xcolor}
\definecolor{webgreen}{rgb}{0,.5,0}

\def\eg{\textit{e.g.}}
\def\ie{\textit{i.e.}}
\def\etc{\textit{etc.}}
\def\md{\mathbf{d}}
\def\Prob{\mathbb{P}}
\def\R{\mathbb{R}}
\def\ed{($\epsilon$,$\delta$)}
\def\ke{($k$,$\epsilon$)}

\DeclareMathOperator{\Var}{Var}
\DeclareMathOperator{\diam}{diam}
\DeclareMathOperator{\Laplace}{Lap}
\DeclareMathOperator{\link}{link}
\DeclareMathOperator{\risk}{risk}
\DeclareMathOperator{\Lap}{Lap}
\DeclareMathOperator{\Hm}{H}
\DeclareMathOperator{\err}{err}
\DeclareMathOperator{\Err}{\mathbf{err}}
\DeclareMathOperator{\Exp}{Exp}

\newcommand{\comment}[1]{}
\title{{\ke}-Anonymity:\\$k$-Anonymity with $\epsilon$-Differential Privacy}

\author{%
{Naoise Holohan, Spiros Antonatos, Stefano Braghin, and P\'{o}l Mac Aonghusa}
\vspace{1.6mm}\\
\fontsize{10}{10}\selectfont\itshape
IBM Research -- Ireland\\
Dublin, Ireland\\
\fontsize{9}{9}\selectfont\ttfamily\upshape
naoise.holohan@ibm.com\\
\{santonat, stefanob, aonghusa\}@ie.ibm.com
}

\begin{document}

\maketitle

\begin{abstract}
The explosion in volume and variety of data offers enormous potential for research and commercial use.
Increased availability of personal data is of particular interest in enabling highly customised services tuned to individual needs.
Preserving the privacy of individuals against reidentification attacks in this fast-moving ecosystem poses significant challenges for a one-size fits all approach to anonymisation.

In this paper we present {\ke}-anonymisation, an approach that combines the $k$-anonymisation and $\epsilon$-differential privacy models into a single coherent framework, providing privacy guarantees at least as strong as those offered by the individual models.
Linking risks of less than 5\% are observed in experimental results, even with modest values of $k$ and $\epsilon$.

Our approach is shown to address well-known limitations of $k$-anonymity and $\epsilon$-differential privacy and is validated in an extensive experimental campaign using openly available datasets.
\end{abstract}



\section{Introduction} \label{sc:intro}

In today's data-driven world, protecting the privacy of individuals' information is of the utmost importance to data curators, both as an ethical consideration and as a legal requirement.
The data collected by governments and corporations potentially holds great value for scientific research and commercial purposes, but to unleash its full potential it must be possible to share data.
Sharing or publishing data while simultaneously preserving user privacy has attracted the attention of researchers in the computer science, mathematics and legal fields for decades~\cite{AW89}.

Data anonymisation is a popular means by which to achieve privacy-preservation in datasets for publication.
Diverse approaches have evolved to deal with this challenge.
In particular, $k$-anonymisation and $\epsilon$-differential privacy are of practical interest and in general use.
The $k$-anonymity framework anonymises data by generalising quasi identifiers, ensuring that an individuals' data is indistinguishable from at least $(k-1)$ others'~\cite{Swe02}.
Differential privacy also preserves privacy by rendering individuals' information indistinguishable~\cite{Dwo06}.
However, instead of using $k$-anonymity's deterministic approach to indistinguishability, differential privacy invokes stochastic indistinguishability by adding noise or perturbing values.

Both $k$-anonymity and $\epsilon$-differential privacy suffer from a number of drawbacks.
In particular, the \emph{curse of dimensionality} of adding extra quasi identifiers to the $k$-anonymity framework results in greater information loss~\cite{Agg05}.
On the other hand, differential privacy has long been criticised for the large information loss imposed on records~\cite{BMS14}.
We show how to overcome these drawbacks by combining $k$-anonymity and $\epsilon$-differential privacy, while simultaneously benefitting from their advantages.

In this paper we present the {\ke}-anonymity framework which augments the deterministic indistinguishability of $k$-anonymity with the stochastic indistinguishability of $\epsilon$-differential 
privacy to produce anonymous datasets.
The curse of dimensionality of $k$-anonymity is addressed by invoking $k$-anonymity only on some quasi identifiers, and imposing $\epsilon$-differential privacy on remaining quasis.
Additionally, we deal with the information loss drawback of differential privacy by leveraging the clusters produced by $k$-anonymity, consequently reducing the noise we are required to add to data.

While both techniques have limitations that can be improved upon in their combination, careful attention must be given to ensure those limitations are not exacerbated further.
In evaluating such a concern, we examine the risk of our approach and establish a metric of record linkage.
Alongside calculating the model's information loss, we can then form an extensive picture of the model's effectiveness.

The structure of the paper is as follows.
In Section~\ref{sc:bg} we outline previous work related to the present paper and also detail the background of $k$-anonymity and $\epsilon$-differential privacy.
In Section~\ref{sc:motivation} we describe the motivating factors in developing the {\ke}-anonymity framework, and give a description of the framework in Section~\ref{sc:hybrid}.
The various means by which we evaluate {\ke}-anonymity is given in Section~\ref{sc:eval}.
Section~\ref{sc:setup} gives details on the experimental set-up of this paper, and experimental results follow in Section~\ref{sc:experiments}.
Concluding remarks are given in Section~\ref{sc:conc}.

\section{Background} \label{sc:bg}

\subsection{Related Work}

Since Sweeney et.\ al.~\cite{Swe02} proposed the $k$-anonymity model, numerous extensions to that model have been proposed to address its weaknesses.
As an example,~\cite{LLV07} proposes $\ell$-diversity and~\cite{LLV07} proposes $t$-closeness, both of which protect against inference-based attacks.
None of the proposed models consider the combination of other anonymisation protocols.

Zakerzadeh et.\ al.~\cite{Zakerzadeh2014TowardsBT} propose to tackle the challenges of high-dimensional privacy by leveraging inter-attribute correlations.
Their approach applies to the $k$-anonymity, $\ell$-diversity, and $t$-closeness models.
Our work applies to all datasets, even in the absence of attribute correlations.

In~\cite{LQS11} the authors propose a way to achieve differential privacy and $k$-anonymity on the same data release.
In order to add a stochastic element to $k$-anonymity (a deterministic system), the authors propose taking a sample at random from the original dataset and then processing this sample to achieve $k$-anonymity prior to publication.
The uncertainty introduced by randomly selecting users to be present in the released (but anonymised) dataset ensures differential privacy can also be achieved.

In contrast,~\cite{KP13} addresses the issue of one-time publishing of non-overlapping counts with $\epsilon$-differential privacy by defining GS, a method that pre-processes the counts by grouping and smoothing them via averaging, and then treating the counts with a differential privacy algorithm.

In~\cite{DS15} the authors present a comparison between the privacy guarantees provided by applying $k$-anonymity and $\epsilon$-differential privacy.
Also, the authors provide a mechanism to approximate the equivalent $\epsilon$ parameter of a $t$-closeness setting and vice-versa.

Recently, Apple has implemented differential privacy to privately collect data from its users~\cite{Gre16b}.
Analysis of its algorithms have found that Apple uses $\epsilon$ values of \num{6} and \num{10}, and values even as high as \num{43} in testing~\cite{TKB17}.

\subsection{Notation}

We attempt to make use of standard mathematical notation throughout the paper.
Scalars are written in regular font (\ie\ $v \in \R$) and vectors are written in boldface (\ie\ $\md \in \R^n$).
Given an integer $m \in \mathbb{Z}$, we let $[m] = [1, m] \cap \mathbb{Z}$.
We let $(\Omega, \Prob, \cal{F})$ be a probability space.
Given a random variable $X: \Omega \to \R$, we write $X(\omega)^n \in \R^n$ as shorthand notation for an $n$-dimensional vector of independent samples of $X$.

\subsection{\texorpdfstring{$k$}{k}-Anonymity}

Usually applied to the release of tabular data consisting of records (rows) and attributes (columns), $k$-anonymity seeks to inhibit the linking of the dataset with external, publicly-available, non-sensitive datasets.
Under $k$-anonymity, each attribute is uniquely labelled as either (i) an explicit identifier, (ii) a quasi identifier, or (iii) a sensitive attribute.
Explicit identifiers, those which uniquely identify the record as an individual (\eg\ name, social security number, address, telephone number, email address, \etc) are removed or masked to prevent reidentification.
Sensitive attributes generally represent the valuable asset of the data, and are usually preserved as is, however, they do pose a risk through inference attacks, and can also serve other purposes for the data curator, such as risk\slash utility based exploration analysis.

In order to prevent linkage attacks with external data, $k$-anonymity requires the combination of all quasi identifiers of an individual to be indistinguishable from the quasi identifiers of at least $(k-1)$ other individuals.
This indistinguishability is achieved by reducing the specificity of the quasi identifiers, commonly referred to as generalisation.
This operation partitions the data into \emph{equivalence classes} of identical (generalised) quasi identifiers, each consisting of at least $k$ individuals.
In cases where $k$-anonymisation cannot be achieved without significant generalisation, suppression of records can be enforced to preserve data utility.

Attempts have been made in the past to correct flaws in the $k$-anonymity framework~\cite{MKG07,LLV07}, but a number of fundamental drawbacks remain.
Firstly, many algorithms for achieving $k$-anonymous datasets are designed to cope best with either categorical or numerical attributes.
For example, Optimal Lattice Optimisation (OLA)~\cite{ola} uses a pre-defined hierarchy of each attribute to arrive at the optimal generalisation; examples of such hierarchies are given in Table~\ref{tbl:kquasis}.
This is primarily geared towards categorical attributes, although other work has examined generalisation hierarchies for numerical attributes~\cite{CCT11}.
In contrast, Mondrian~\cite{mondrian} performs on-the-fly bisections on the entire attribute state space to optimally split the dataset into equivalence classes of size at least $k$.
This approach works best with numerical attributes, but its application to categorical data has previously been considered~\cite{Tor04}.

\begin{table}[h]
\centering
	\caption{Examples of Generalisation Hierarchies (* denotes omitted details)}
	\label{tbl:kquasis}
        \scriptsize
	\begin{tabular}{|l|l|l|}
		\hline
	{\bf Attribute} & {\bf \# Levels} & {\bf Description} \\
\hline
	Year of Birth & 5 & Value, 2-yr interval, 4-yr interval, 8-yr interval, * \\
	Gender & 2 & Value, Person \\
	Race & 2 & Value, * \\
	Marital Status & 3 & Value, Alone\slash In marriage, *\\
	ZIP code & 6 & Value, XXXX*, XXX**, XX***, X****, ***** \\
\hline
	\end{tabular}
\end{table}

Additionally, $k$-anonymisation has traditionally only been used with deterministic tools, meaning the output of the algorithm will always be identical.
This means an attacker will also know for certain where a particular user lies in the anonymised dataset, if the attacker knows the user is present in the dataset.
The lack of any uncertainty in the output of the dataset strengthens the hand of an attacker.
By blending $k$-anonymisation with differential privacy, we can add an element of non-determinism.

\subsection{\texorpdfstring{$\epsilon$-}{}Differential Privacy} \label{sc:bg:dp}

At the same time as $k$-anonymity was being developed, other researchers were independently inventing a new way to achieve privacy-preserving data releases.
Instead of considering privacy as a function of the dataset, the researchers proposed that privacy should be a function of the mechanism.
Therefore, any dataset which the mechanism processes would be guaranteed the same level of privacy as any other dataset, no matter how extreme or sensitive the dataset was.

The idea of stochastic indistinguishability for data privacy was proposed in~\cite{DN03}, and eventually culminated in the formal proposal of \emph{differential privacy} by Cynthia Dwork~\cite{Dwo06}.
When applied in the context of privacy-preserving data mining (PPDM), as envisaged by Dwork, differential privacy enforces that query answers from similar datasets be statistically indistinguishable, as determined by the privacy parameter $\epsilon$.
This statistical indistinguishability ensures that the differential in information gain between two similar datasets is capped, hence giving rise to the name \emph{differential} privacy.

\subsubsection{Model Considerations}
Although it was initially envisaged for PPDM, differential privacy can also be applied to privacy-preserving data publishing (PPDP) applications.
In this scenario, a data curator\slash owner publishes a dataset which conforms to differential privacy, with which researchers and other interest parties can then analyse and\slash or ask queries.
The guarantees provided by differential privacy ensure that all subsequent queries asked will also satisfy differential privacy~\cite[Theorem~4]{HLM15}.
This also allows for the asking of an unlimited number of queries while still satisfying differential privacy, a situation which is not trivial to achieve in PPDM.

One instantiation of PPDP is that of randomised response (RR)~\cite{War65}.
Under randomised response, the privacy-preserving action is performed before the data is collected by the curator (\eg\ a respondent using a coin-flip to decide whether to answer a survey question truthfully or not).
Randomised response is actively used when conducting surveys of a sensitive nature~\cite{FL88,UPC17}, as the privacy of the respondents is protected even from the surveyors.
Randomised response can also be applied to datasets before publication.

In this context, randomised response corresponds to any PPDP protocol which is independently applied to each row of a dataset.
We can therefore achieve statistical indistinguishability between individual records in a dataset, thereby protecting the privacy of the individuals involved.
This is similar to the approach taken by $k$-anonymity, where deterministic indistinguishability is used to protect the privacy of individuals.
However, while the approach to protect privacy by indistinguishability is shared with $k$-anonymity, using differential privacy to achieve statistical indistinguishability provides strong guarantees against adversary attacks~\cite{KS08}.
Satisfying differential privacy in randomised response mechanisms has been studied in the past for simple mechanisms~\cite{HLM17}

\subsubsection{Definition \& Parameters}
Before defining differential privacy, we first present the following objects: a record-space $D$ (where individual records belong), a dataset size $n \in \mathbb{N}$, a query output space $E_Q$ (which may be numerical, categorical, higher-dimensional, \etc) and a probability space $(\Omega, \mathcal{F}, \Prob)$.
In defining differential privacy, we represent a query $Q: D^n \to E_Q$ operating on a database $\md \in D^n$ as a random variable $X_{Q, \md}: \Omega \to E_Q$, which we call the \emph{response mechanism}.
Then, taking a privacy parameter $\epsilon \ge 0$, the response mechanism $X_{Q, \md}$ satisfies $\epsilon$-differential privacy if
\begin{equation}\label{eq:dp}
\Prob(X_{Q, \md} \in A) \le e^\epsilon \Prob(X_{Q, \md'} \in A),
\end{equation}
where the datasets $\md$ and $\md'$ differ in exactly one record (\ie\ one individual), and where $A \subseteq E_Q$ is measurable.

The concept of \emph{relaxed differential privacy} was later proposed by~\cite{KS08} to relax the strict condition of differential privacy.
Given a relaxation parameter $\delta \in [0,1]$, a mechanism $X_{Q, \md}$ is said to satisfy {\ed}-differential privacy if
\begin{equation}\label{eq:relaxeddp}
\Prob(X_{Q, \md} \in A) \le e^\epsilon \Prob(X_{Q, \md'} \in A) + \delta,
\end{equation}
where the datasets $\md$ and $\md'$ differ in exactly one record (\ie\ one individual), and where $A \subseteq E_Q$ is measurable.
For simplicity, only $\epsilon$-differential privacy is considered in this paper, but we note that all the results and methods can be applied to relaxed differential privacy simply by ensuring that the differential privacy mechanism satisfies (\ref{eq:relaxeddp}).

It can be seen that the definition of differential privacy is independent of the type and distribution of the data being considered.
This follows the underlying principle of requiring privacy to be a function of the mechanism and not the data.
Under differential privacy, two datasets can meet the same privacy guarantee (\ie\ $\epsilon$-differential privacy or {\ed}-differential privacy) independently of whom the data concerns.

\subsubsection{Response Mechanisms}\label{sc:bg:dp:rm}
The data-independence of the differential privacy model becomes apparent when considering its application to specific data types.
In this paper we focus on the application of differential privacy to a single column of numeric data, and the study of this application is well-developed in the literature.
The \emph{Laplace mechanism} is one popular such mechanism~\cite{Dwo06,HLM16}, having first been proposed by Dwork.

We denote by $\Lap(\mu, b)$ the Laplace distribution with mean $\mu$ and variance $2b^2$, whose PDF $f_{\Lap(\mu, b)}: \R \to \R_{\ge0}$ is given by
$$f_{\Lap(\mu, b)}(x) = \frac{1}{2b} e^{-\frac{|x-\mu|}{b}}.$$
The diameter $\diam: \R^n \to \R_{\ge0}$ of a dataset $\md$ is given by its range of values, \ie
\begin{equation}\label{eq:diam}
\diam(\md) = \max_i(d_i) - \min_i(d_i).
\end{equation}
Then, for $\md \in \R^n$ and a random variable $Y \sim \Lap\big(0, \frac{\diam(\md)}{\epsilon}\big)$, it is shown in~\cite{HLM16} that the response mechanism $X_{\md}: \Omega \to \R^n$ given by
$$X_\md(\omega) = \md + Y(\omega)^n$$
satisfies $\epsilon$-differential privacy.

Although the definition of differential privacy is independent of the data, we note that the definition of the response mechanism does depend on the data, by virtue of the presence of $\diam(\md)$.
In effect, the larger the range of data in the dataset, the more noise must be added to preserve privacy (to make individuals indistinguishable).
It is also worth noting that a dataset without any variation (\ie\ $\diam(\md)=0$) requires no noise addition, as individuals in such a dataset are already indistinguishable.

\section{Motivation}\label{sc:motivation}

Traditionally, when it comes to privacy-preserving data publishing, data owners employ a variety of techniques and algorithms to achieve an anonymous result.
Data masking techniques are invoked to remove the obvious explicit identifiers, like names, addresses, phone numbers \etc\
In order to achieve stronger privacy guarantees, anonymisation algorithms are used to protect the quasi identifiers; non-explicit identifiers that can be used in combination to uniquely identify individuals.
Widely used approaches include performing $k$-anonymity on a set of quasi identifiers or applying $\epsilon$-differential privacy to the entire dataset. 

Our work has been motivated by the following use case.
Medical data sharing often involves dataset that includes both categorical and numerical attributes.
For example, a medical dataset might contain demographic information, such as gender and age, disease information, like diagnosis codes, and measurements (for example height, weight and blood readings).
In the majority of cases, data owners select a single anonymisation algorithm to apply. 
By applying a single anonymisation approach, the final result might suffer from high information loss due to the very nature of the applied algorithms.
For example, applying $k$-anonymity on a high-dimensional dataset results in large information loss, even for small values of $k$.
The main driving force of our work is that the combination of approaches yields better results than applying a single algorithm.  

The {\ke}-anonymity approach proposed in this paper first performs $k$-anonymisation on a subset of the quasi identifiers and then $\epsilon$-differential privacy on the remaining quasi identifiers with different settings for each equivalence class of the $k$-anonymous dataset.  
In this section we describe how our approach compares with existing widely-used approaches and demonstrate why {\ke}-anonymity has obvious benefits over applying single, isolated approaches. 

\subsection{The Curse of Dimensionality for \texorpdfstring{$k$}{k}-anonymity}

The addition of quasi identifiers to the $k$-anonymity model results in greater generalisation and information loss on those quasi identifiers.
If we seek to reduce information loss on any particular quasi identifiers, we must reduce the $k$-value which results in a corresponding increase in  reidentification risk, or remove some quasi identifiers from the data completely.
This poses a problem with data that contains many quasi identifiers that are valuable to the data for analytical purposes.
This phenomenon has previously been studied in~\cite{Agg05}.

We denote as \emph{information loss} the quantification of the distortion imposed on data as part of the anonymisation procedure.
Calculating information loss on an anonymised dataset is of the utmost importance, not least in being able to select the optimal solution.
Selecting which information loss metric to use depends on a host of factors, including the type of data involved, the algorithm used to achieve privacy and the possible end-use of the data.

For $k$-anonymity, when considering information loss on generalised attributes, we consider two metrics.
\begin{enumerate}
\item For attributes generalised according to a hierarchy, we consider \emph{categorical precision}.
Given a hierarchy with $h$ levels, a value which is generalised to a level $l$ (where $l=0$ denotes no generalisation and $l=h-1$ denotes full generalisation) has information loss of $\frac{l}{h-1}$ according to categorical precision.
For example, given \emph{ZIP code} as an attribute ($h=6$ according to Table~\ref{tbl:kquasis}), a generalised value of 237** has information loss of $\frac{2}{5}=0.4$ according to categorical precision.
\item For numerical attributes generalised without a hierarchy, we consider \emph{numerical precision}.
The information loss of such an attribute is given as the ratio of the range of the generalised interval to the range of the dataset.
For example, given a dataset with \emph{height} in the range $[150, 190]$ and a generalised value of $[165-180]$, the information loss according to numerical precision is $\frac{180-165}{190-150}=\frac{15}{40}=0.375$.
\end{enumerate}

\begin{figure*}[tb]
	\includegraphics[width=\textwidth]{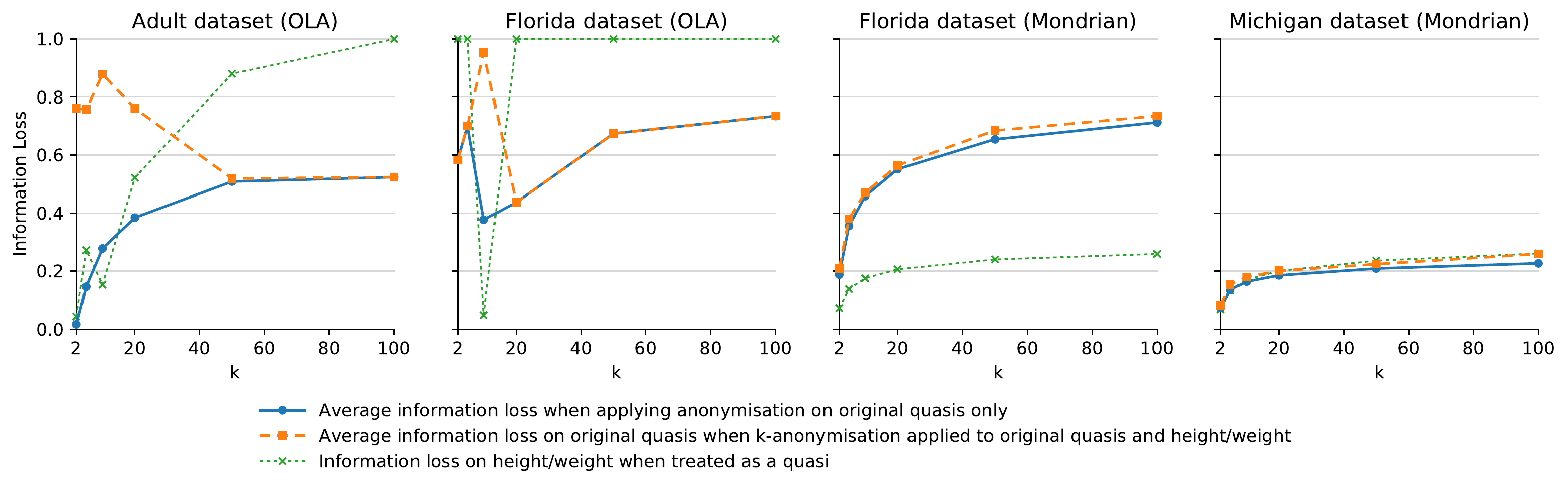}
  \caption{\label{fig:mv:dimensions} Curse of dimensionality illustrated by comparing information loss on original quasi identifiers, before and after the consideration of height\slash weight as quasi identifiers.}
\end{figure*}

This curse of dimensionality is illustrated in Figure~\ref{fig:mv:dimensions}. We applied both OLA and Mondrian $k$-anonymity 
algorithms on three real datasets that were augmented to include height and weight 
attributes (more details of the datasets and their properties used in this example are given in Section~\ref{sc:setup}).
The information loss on each dataset's quasi identifiers is shown before and after the addition of an extra quasi identifier 
(namely a \emph{height} attribute for the Adult and Florida datasets, and a \emph{weight} attribute for the Michigan dataset).
Information loss on height\slash weight is measured by categorical precision in OLA, and numerical precision in Mondrian.

We observe that in the case of OLA, adding the height as a quasi identifier the impact on the information loss is significant. We can 
spot two major patterns. The first one is that for small $k$ values adding height as a quasi leads to much higher information loss of the original quasis.
For example, in the Adult dataset for $k$ equal to 20, the information loss was doubled to 0.8 up from 0.4. 
The second is that for larger values of $k$ (more than 20), the information loss for the original quasis remains the same but the loss for the height
attribute is more than 0.8. We therefore have the quasi identifiers experiencing additional generalisation at the expense of adding an extra quasi, which itself 
has been excessively generalised in order to be included. In practical terms, the height attribute was transformed close to its maximum generalisation level. 
For the case of Mondrian, we observe that our approach results to a relative 4-10\% decrease on the information loss of the original quasis while 
presenting the same or less relative average error on the height/weight attributes.  

\subsection{Global Differential Privacy versus \texorpdfstring{\ke}{(k,e)}-Anonymity}

The diversity of the data being processed by a differential privacy response mechanism has an impact on the level of noise added, with greater diversity requiring greater perturbation.
In numerical attributes, the diversity is measured as the diameter of the data.

By first applying $k$-anonymity, we can take advantage of the partitioning into equivalence classes for applying differential privacy.
If we consider the addition of an extra quasi identifier to a $k$-anonymous dataset, the noise addition can be reduced by considering each equivalence class to be a separate dataset.
We then reduce the quantity of noise required to achieve $\epsilon$-differential privacy, while maintaining the indistinguishability guarantee of differential privacy.

Formally, let's consider the case of adding Laplace-distributed noise to achieve differential privacy (see Section~\ref{sc:bg:dp:rm}).
The magnitude (as determined by the scale factor) of the Laplace noise required to achieve $\epsilon$-differential privacy is given as $\frac{\diam}{\epsilon}$.
Given a dataset $\md \in \R^n$, and an equivalence class of size $m$ as a subset of that dataset $\mathbf{ec} \in \R^m$, 
where, for each $i \in [m]$, there exists a unique and distinct $j \in [n]$ such that $ec_i = d_j$, 
we note that
$$\diam(\md) \ge \diam(\mathbf{ec}),$$
since $\min(\mathbf{ec}) \ge \min(\md)$ and $\max(\mathbf{ec}) \le \max(\md)$.
Hence the variance $\Var$ of the noise being added satisfies
\begin{align*}
\Var\left(\Lap\left(0, \frac{\diam(\md)}{\epsilon}\right)\right) &= 2\left(\frac{\diam(\md)}{\epsilon}\right)^2\\
&\ge \Var\left(\Lap\left(0, \frac{\diam(\mathbf{ec})}{\epsilon}\right)\right).
\end{align*}

\begin{figure}[tb]
	\centering
	\includegraphics[width=0.75\columnwidth]{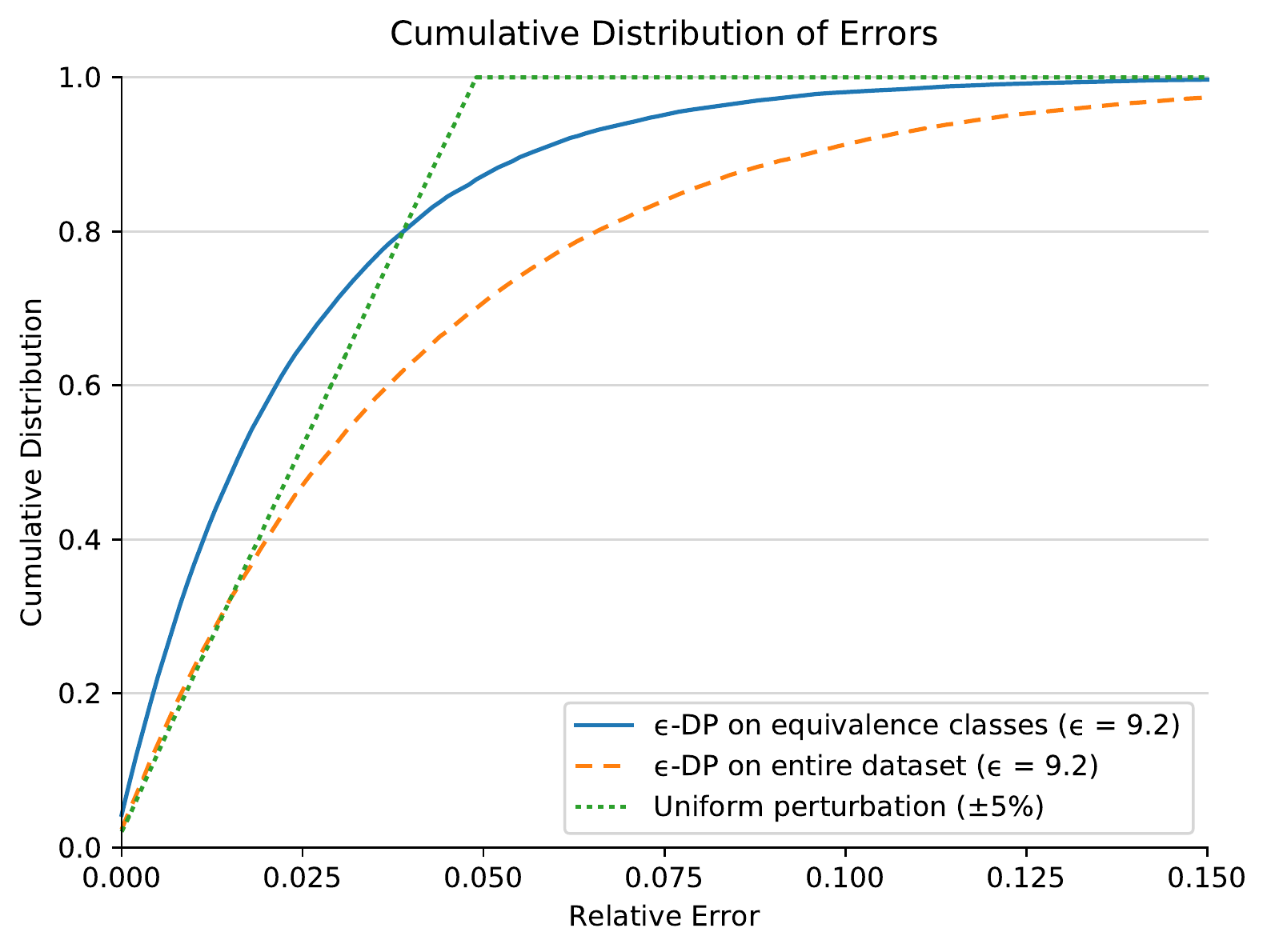}
  \caption{\label{fig:mv:iloss} Distribution of relative errors when applying differential privacy before and after $k$-anonymisation, and when fixed perturbation is applied.}
\end{figure}

Figure~\ref{fig:mv:iloss} shows the cumulative relative error distribution of applying differential privacy to the entire dataset alongside that of applying differential privacy 
on equivalence classes. We observe that the relative error when applying differential privacy to the entire dataset is consistently larger than our approach.

\subsection{Uniform Additive Perturbation versus \texorpdfstring{\ke}{(k,e)}-Anonymity}

It is common to use uniform noise in additive perturbation mechanisms for data privacy purposes~\cite{KDW03,AW89}.
While it is possible to obscure a value by using uniform noise, it is not guaranteed that such a technique will ensure indistinguishability between values.
Due to the finite support of the uniform distribution, there will be output values possible which will map directly back to its original.

For example, using additive perturbation with $\pm10\%$ uniformly distributed noise, it is not possible to guarantee the indistinguishability of the values $1$ and $2$, as $1$ will be perturbed somewhere in the range $[0.9, 1.1]$, and $2$ will be perturbed in $[1.8, 2.2]$.
Similarly, with noise of $\pm100\%$, the value $2$ will be distinguishable from $1$ if a positive perturbation is applied.

The privacy guarantees provided by our approach far outweigh those provided by uniform perturbation.
Nevertheless, we can easily match the information loss of a differential privacy mechanism with the same information loss incurred by a uniform perturbation.
Figure~\ref{fig:mv:iloss} shows the cumulative relative error distribution of a uniform perturbation ($\pm 5\%$, dotted line) alongside that of a differential privacy mechanism with the same mean error (solid line).
Similarly, given the mean error from the differential privacy mechanism, we can represent it in the form of the error of a uniform perturbation mechanism for ease of understanding.
Information loss and error is further discussed in Section~\ref{sc:eval:iloss}.

\section{\texorpdfstring{\ke}{(k,e)}-Anonymity}\label{sc:hybrid}

We now detail the formulation of the {\ke}-anonymity privacy framework.
An outline of this workflow is given in Algorithm~\ref{alg:workflow}.

As a first step, we classify the attributes into \emph{explicit identifiers}, \emph{quasi identifiers} and \emph{sensitive attributes}.
We then further partition the quasi identifiers into $k$-quasis, upon which we apply traditional $k$-anonymity, and $\epsilon$-quasis upon which we apply $\epsilon$-differential privacy.
The selection of the $k$-quasis and the $\epsilon$-quasis can be done based on the use case of the data, the clustering algorithm being used for $k$-anonymity, or any other reason deemed appropriate by the data curator.
An example of attribute classification is detailed in Table~\ref{tbl:attrclass}.
We selected the numerical attributes as $\epsilon$-quasis inkeeping with the focus of this paper.

\begin{table}[ht]
\centering
	\caption{Attribute classification example}
	\label{tbl:attrclass}
        \scriptsize
	\begin{tabular}{|l|l|}
			\hline
		Attributes: & Name, SSN, Gender, Year of Birth, ZIP code,\\
		& Marital status, Height, Weight\\
	        \hline \hline
        Explicit Identifiers: & Name, SSN \\
    	    \hline
        Sensitive attributes: & Blood pressure\\
        	\hline
        Quasi identifiers: & Gender, Year of Birth, ZIP code, Marital Status,\\
        & Height, Weight\\
	        \hline \hline
        $k$-Quasis: & Gender, Year of Birth, ZIP code, Marital Status\\
    	    \hline
        $\epsilon$-Quasis: & Height, Weight \\
        	\hline
	\end{tabular}
\end{table}

The $k$-quasis are then processed to satisfy $k$-anonymity, thereby clustering the dataset into equivalence classes of at least $k$ records.
Certain $k$-anonymisation algorithms, including OLA, support suppression of records in order to reduce the generalisation required to achieve $k$-anoymity.

The next step is to apply differential privacy to the $\epsilon$-quasis.
We only take inspiration from the general concept of differential privacy and its notion of stochastic indistinguishability, rather than applying differential privacy in its original formulation.
While differential privacy originally sought to make an individual in a dataset indistinguishable from any other individual who could possibly be in the dataset, our approach is intricately linked to our threat model.

As $k$-anonymisation has already taken place, an attacker can therefore always uniquely identify the equivalence class to which an individual belongs by examining the $k$-quasis.
We must therefore ensure reidentification cannot occur using the $\epsilon$-quasis.

By making the $\epsilon$-quasis stochastically indistinguishable, we can prevent against such reidentification attacks.
However, as an attacker knows which equivalence class to which an individual belongs, we need only make the $\epsilon$-quasis indistinguishable within each equivalence class.
$\epsilon$-Quasis need not be indistinguishable between separate equivalence classes, since our attack model ensures there will never be a reidentification risk between equivalence classes.

In a differential privacy context, we can therefore treat each equivalence class as an independent population of individuals.
For example, for a `height' $\epsilon$-quasi and an equivalence class with a maximum `height' of \num{170}cm, we need not make this $\epsilon$-quasi in this equivalence class indistinguishable with a `height' value of \num{180}cm, since it doesn't occur in the equivalence class population.
By reducing the range of values of the population, we reduce the magnitude of noise required to achieve differential privacy, thereby reducing the distortion imposed on the data.

In the application of differential privacy, it is important that an equivalence class is considered a collection of records, not an independent collection of attributes.
If each $\epsilon$-quasi is processed independently, correlation between attributes will result in additional privacy leakage.
This leakage can be demonstrated by copying the same $\epsilon$-quasi multiple times in a dataset.
If differential privacy is applied independently to each of the copies, an adversary can average the perturbed copies to obtain an unbiased estimate of the original value.
Such an attack would not be possible if differential privacy was instead applied to each record.

Finally, we shuffle the records to prevent against order-based attacks.

\begin{algorithm}[tb]
  \KwIn{Dataset $D$}
  \KwOut{{\ke}-anonymous dataset $D^{\prime}$}
  
  Perform \emph{attribute classification}\;
  kQID = $k$-quasis\;
  eQID = $\epsilon$-quasis\;
  Remove explicit identifiers\;
  Perform $k$-anonymisation on kQID\;
  \ForEach{Equivalence class $ec$ of the $k$-anonymous dataset}{%
    Apply $\epsilon$-differential privacy mechanism to eQID to the record of $ec$\;
  }
  Merge the $\epsilon$-differentially private and $k$-anonymous equivalence classes\;
  Merge the $\epsilon$-differentially private and $k$-anonymous equivalence classes\;
  Shuffle records\;
  \caption{The \ke{}-Anonymity Workflow}
  \label{alg:workflow}
  \end{algorithm}

\section{Model Evaluation} \label{sc:eval}

In this section we detail the methods and metrics by which we evaluate the {\ke}-anonymity model.
Since we introduce a new anonymity model, we need to provide the framework to measure both information loss
and re-identification risk. We present the threat model under which we consider risk as well as our approach to alleviate
the risk introduced by the stochastic indistinguishability of $\epsilon$-differential privacy.

\subsection{Information Loss}\label{sc:eval:iloss}

In this paper we only consider numerical attributes as $\epsilon$-quasis and for these, we consider the \emph{average relative error} as a measure of information loss.
As we are dealing with random variables, we consider the expected error that will result on a value from applying the mechanism.
We can then calculate the total error on the dataset as the average of errors on each value.

In general, the relative error of an anonymised value $v^\prime$, given an original value of $v \in \R$ is given by
$$\err(v^\prime) = \left|\frac{v^\prime - v}{v}\right|.$$

Formally, given a value $v \in \R$, and a noise random variable $Z_{\epsilon, \diam}: \Omega \to \R$, we calculate the error $\err: \R \to \R_{\ge 0}$ by
\begin{align}
\err(v + Z_{\epsilon, \diam}(\omega)) &= \mathbb{E}\left[\left|\frac{Z_{\epsilon, \diam}(\omega)}{v}\right|\right]\\
&=\frac{\mathbb{E}\left[|Z_{\epsilon, \diam}(\omega)|\right]}{|v|}
\end{align}
For a set of $n$ values, $\md \in \R^n$ (\ie\ the $\epsilon$-quasi column of the dataset), the total error $\Err: \R^n \to \R_{\ge 0}$ is given by
$$\Err\left(\md + Z_{\epsilon, \diam}(\omega)^n\right) = \frac{1}{n} \sum_{i = 1}^n \err\left(d_i + Z_{\epsilon, \diam}(\omega)\right).$$

As we apply the differential privacy mechanism to each equivalence class independently, we calculate the error per equivalence class.
Given an equivalence class of size $m$, denoted by $\mathbf{ec} \in \R^m$, we can calculate its information loss as a function of its harmonic mean.

The harmonic mean $\Hm$ of $\mathbf{ec}$ is given by
$$\Hm(\mathbf{ec}) = \frac{m}{\sum_{i=1}^m \frac{1}{ec_i}},$$
which allows us to calculate the information loss as follows:
\begin{align}
\Err(\mathbf{ec}+(Z_{\epsilon, \diam(\mathbf{ec})})^m) &= \frac{1}{m} \sum_{i=1}^m \frac{\mathbb{E}[|Z_{\epsilon, \diam(\mathbf{ec})}|]}{|ec_i|}\nonumber\\
 &= \mathbb{E}[|Z_{\epsilon, \diam(\mathbf{ec})}|] \frac{1}{m} \sum_{i=1}^m \frac{1}{|ec_i|}\nonumber\\
 &= \mathbb{E}[|Z_{\epsilon, \diam(\mathbf{ec})}|] \frac{1}{\Hm(\mathbf{ec})}\label{eq:harmoniciloss}
\end{align}

We can further simplify (\ref{eq:harmoniciloss}) in the case where the noise we add is drawn from a Laplace distribution.
As noted in Section~\ref{sc:bg:dp:rm}, the addition of Laplace-distributed noise with mean zero and variance $2b^2$ (where $b=\frac{\diam}{\epsilon}$) satisfies $\epsilon$-differential privacy.
We therefore have $Z_{\epsilon, \diam} \sim \Lap\left(0, \frac{\diam}{\epsilon}\right)$, giving
\begin{align*}
\mathbb{E}[|Z_{\epsilon, \diam}|] &= \mathbb{E}\left[\left|\Lap\left(0, \frac{\diam}{\epsilon}\right)\right|\right]\\
&= \mathbb{E}\left[\Exp\left(\frac{\epsilon}{\diam}\right)\right]\\
&= \frac{\diam}{\epsilon},
\end{align*}
where $\Exp(\lambda)$ denotes the negative exponential distribution with mean $\frac{1}{\lambda}$, whose PDF $f_{\Exp(\lambda)}: \mathbb{R}_{\ge 0} \to \mathbb{R}$ is given by
$$f_{\Exp(\lambda)}(x) = \lambda e^{-\lambda x}.$$
It can be shown for a random variable $X \sim \Laplace(\mu, b)$, that $|X-\mu| \sim \Exp\left(\frac{1}{b}\right)$.

We therefore have the information loss for an equivalence class $\mathbf{ec}$:
\begin{equation}\label{eq:ecerror}
\Err(\mathbf{ec}+(Z_{\epsilon, \diam(\mathbf{ec})})^m) = \frac{\diam(\mathbf{ec})}{\epsilon \Hm(\mathbf{ec})}.
\end{equation}

To calculate the information loss for an entire dataset, we average over the weighted sums of the equivalence classes:
\begin{equation}\label{eq:dberror}
\Err(\md + (Z_{\epsilon, \diam})^m) = \frac{1}{\epsilon} \sum_{\mathbf{ec}} \frac{\diam(\mathbf{ec})}{\Hm(\mathbf{ec})} \frac{|\mathbf{ec}|}{n},
\end{equation}
where $|\mathbf{ec}| \in \mathbb{N}$ denotes the number of records in $\mathbf{ec}$.

Using (\ref{eq:dberror}), the error of a dataset can be determined before applying differential privacy.
Depending on the use case, this error calculation can be used in the determination of an $\epsilon$-value to preserve the utility of the data as required.

\subsection{Linking Risk}\label{sc:eval:link}

Before proceeding with the formulation of the linking risk, it is important to describe the considered threat model under which our model operates.
We adopt a worse-case threat model for the purpose of this paper.
We are seeking to prevent reidentification attacks by an attacker with access to the (unaltered) explicit identifiers and quasi identifiers of the dataset.
We assume the attacker has no knowledge of the sensitive attribute of the dataset on any individual.
This is a reasonable assumption to make from a linking attack perspective, since the sensitive attributes of a dataset frequently represent the value of a dataset, and hence are new information that was not previously available.
In reidentifying a record of a dataset, an adversary would attempt to link the quasi identifiers of an individual from their unanonymised copy with the anonymised dataset.
The challenge is to ensure sufficient indistinguishability in the anonymised dataset (using a combination of deterministic and stochastic indistinguishability) to prevent reidentification, while also preserving the utility of the dataset.

For analysing the risk of a {\ke}-anonymous dataset, we present the following simple, nearest-neighbour, linking model.
Given a {\ke}-anonymous dataset and its equivalence classes, we seek to link records back to the records of the original dataset.

For each record in the anonymised dataset, we select a candidate record in the original dataset which lies in the same equivalence class as the one being investigated.
This candidate is selected as being the one whose $\epsilon$-quasi (in the original dataset) lies closest to the anonymised $\epsilon$-quasi of the record being investigated.
If the investigated record and the candidate record correspond to the same record, then we say a successful link has occurred.
We repeat this process for every record in the anonymised dataset, and report the \emph{linking risk} as the fraction of records which have been successfully linked in the manner.

Formally, given an equivalence class $\mathbf{ec} \in \R^m$, and its anonymised version $\mathbf{ec}^\prime = \mathbf{ec} + \Lap\left(0, \frac{\diam(\mathbf{ec})}{\epsilon}\right)^n \in \R^m$, a successful link for record $i \in [m]$ has occurred if $ec^\prime_i$ is closer to $ec_i$ than all other $ec_j$.

We construct an indicator function $\phi_{\mathbf{ec}^\prime}: [m] \to \{0, 1\}$ which signals a successful link:
$$\phi_{\mathbf{ec}^\prime}(i) = \begin{cases} 1, & \text{if } |ec^\prime_i - ec_i| = \min_{j \in [m]} |ec^\prime_i - ec_j|;\\ 0, & \text{otherwise.}\end{cases}$$
We then define the number of linkable records, $\link: \R^n \to [0,m]\cap\mathbb{Z}$, as follows:
$$\link(\mathbf{ec}^\prime) = \sum_{i \in [m]} \phi_{\mathbf{ec}^\prime}(i).$$

For an entire anonymised dataset $\md^\prime \in \R^n$, we calculate its linking risk, $\risk: \R^n \to [0,1]$, as the proportion of records in the dataset that can be successfully linked back to its original record:
$$\risk(\md^\prime) = \frac{1}{n} \sum_{\mathbf{ec}^\prime} \link(\mathbf{ec}^\prime).$$

It is important to note that this does not constitute a reidentification risk, as records in the same equivalence class with the same value for the $\epsilon$-quasi will remain indistinguishable.

\subsection{Confidence-based \texorpdfstring{$k$}{k}-Anonymity}\label{sc:eval:pseudokanon}

Although the implementation of differential privacy makes individual values stochastically indistinguishable, it is possible to obtain a measure of deterministic indistinguishability for an instance of a {\ke}-anonymous dataset.
First, we compute a confidence interval around the differentially private value, based on the known $\epsilon$ and diameter of the $\epsilon$-quasi of its equivalence class.
This gives a confidence interval around from where the original value came, thereby allowing an adversary to link the anonymous record using his\slash her table of un-anonymised quasis.

Given an equivalence class $\mathbf{ec} \in \R^m$, and its anonymised version $\mathbf{ec}^\prime = \mathbf{ec} + \Lap\big(0, \frac{\diam(\mathbf{ec})}{\epsilon}\big)^m$, and a confidence $c \in [0,1]$, we seek to determine a linking range $r_c \in \R_{\ge 0}$ such that for each $i \in [m]$,
\begin{equation}\label{eq:eval:confdef}
\Prob(ec_i \in [ec_i^\prime - r_c, ec_i^\prime + r_c]) = c.
\end{equation}
We can simplify (\ref{eq:eval:confdef}) based on our assumption of Laplace-distributed noise as follows, where $b = \frac{\diam(\mathbf{ec})}{\epsilon}$:
\begin{align*}
\Prob\left(ec_i \in [ec_i^\prime - r_c, ec_i^\prime + r_c]\right) &= \Prob\left(|\Lap(0, b)| \le r_c \right) \\
&= \Prob\left(\Exp\left(\frac{1}{b}\right) \le r_c \right)\\
&= \int_0^{r_c} \frac{1}{b} e^{-\frac{x}{b}} dx\\
&= - \Big[ e^{-\frac{x}{b}} \Big]_0^{r_c}\\
&= 1 - e^{-\frac{r_c}{b}}\\
&= c.
\end{align*}
We then solve for $r_c$, and find
\begin{align*}
r_c &= -b \ln(1-c)\\
&= - \frac{\diam(\mathbf{ec})}{\epsilon} \ln(1-c).
\end{align*}

After completing the {\ke}-anonymity process on a dataset, we use the above confidence interval calculations to evaluate the linking vulnerabilities of the anonymised dataset.
For each anonymised equivalence class $\mathbf{ec}^\prime \in \R^m$, we analyse each record $ec_i^\prime, i \in [m]$ and count the number $\ell_i$ of records in $\mathbf{ec}$ that fall within its $c$-confidence interval:
$$\ell_i = |\{j: ec_j \in [ec_i - r_c, ec_i + r_c]\}|.$$

To maintain the risk profile of $k$-anonymity, we propose $c$-confident $k$-anonymity for $\epsilon$-quasis, whereby records with $0 < \ell_i < k$ are suppressed.
We choose not to suppress records where $\ell_i = 0$, as these are records which the differential privacy process has rendered outlying, thereby posing little reidentification risk.
Furthermore, we propose to suppress entire equivalence classes which are left with fewer than $k$ records after this process, \ie\ equivalence $\mathbf{ec} \in \R^m$ is suppressed whenever
$$|\mathbf{ec}| - |\{i : \ell_i < k\}| < k.$$

\section{Experimental Setup}\label{sc:setup}

In order to validate our model, we used three datasets for experimental analysis.
The three datasets are publicly available, and in order to increase the granularity of the data, we added synthetically-created additional attributes.
We used the training set of the UCI Adult dataset~\cite{Lichman:2013}, and samples of the publicly-available voter lists from Florida~\cite{floridavoters} and Michigan~\cite{michiganvoters} as the base datasets.

For the purpose of this paper, we are limiting the scope of the $\epsilon$-quasis.
We limit our analysis to using only one $\epsilon$-quasi in each dataset, and additionally, require that this $\epsilon$-quasi is a numerical attribute.

As the three datasets in question contained no numerical attribute, we generated a synthetic numerical attribute for each.
This consisted of adding a \textit{height} column to the Adult and Florida datasets and a \textit{weight} column to the Michigan dataset.
In generating the height columns, we generated normal distributions for age and gender, fitting the data provided in~\cite{FGO12} for the US population.
The height for each individual was generated by independently sampling from the distribution corresponding to their age and gender.
Similarly, in generating the weight column, we generated log-normal distributions for age and gender, fitting the data provided in~\cite{FGO12} for the US population.
The weight for each individual was generated by independently sampling from the distribution corresponding to their age and gender.

Table~\ref{tbl:datasets} lists the datasets and their respective attributes as used in our experiments.
Attributes which were synthetically created for analytical purposes are listed in \textit{italic font}.

\begin{table*}[tb]
\centering
\caption{Dataset Attributes (synthetic attributes listed in \textit{italic})}
\label{tbl:datasets}
\scriptsize
\begin{tabular}{|l|l|l|l|}

\hline
 & Adult & Florida & Michigan \\
\hline
	Number of records: & \num{32440} & \num{134791} & \num{72825} \\
	Direct identifiers: & \textit{Patient Id}, \textit{Name}, \textit{E-mail} & Name, Surname, Address, E-mail, Phone Number & Name, Surname, Address \\
	Quasi identifiers: & Year of Birth, Gender, Race, Marital Status, \textit{Height} & ZIP code, Gender, Year of Birth, Race, \textit{Height} & ZIP Code, Gender, Year of Birth, \textit{Weight} \\
\hline \hline
	$k$-Quasis: & Year of Birth, Gender, Race, Marital Status & ZIP code, Gender, Year of Birth, Race & ZIP Code, Gender, Year of Birth \\
	$\epsilon$-Quasis: & \textit{Height} & \textit{Height} & \textit{Weight}\\
	\hline
\end{tabular}
\end{table*}

For each experiment, we first removed all explicit identifiers from the dataset, as listed in Table~\ref{tbl:datasets}.
We then created $k$-anonymous versions of the datasets using both the OLA and Mondrian $k$-anonymisation algorithms.
We ran experiments for $k$ values of \num{2}, \num{5}, \num{10}, \num{20}, \num{50}, and \num{100}.
The generalisation hierarchy used for each of the $k$-quasis (all quasi identifiers except the $\epsilon$-quasis height and weight) is listed in Table~\ref{tbl:kquasis}.
The maximum suppression applied by the OLA algorithm was fixed at \num{5}\%.
Table~\ref{tbl:olasupp} details the suppressions applied by OLA for the three datasets and for the $k$ values of interest.

For the application of differential privacy, experiments were run using $\epsilon$ values of \num{0.05}, \num{0.5}, \num{1}, \num{2}, \num{4}, \num{8} and \num{16}.
In some figures in this paper, for ease of readability, some values of $\epsilon$ are not plotted.
In each case, the reported results are based on an average over 30 runs, given that applying differential privacy is a stochastic process.

\begin{table}[tb]
\centering
	\caption{Effective suppression applied by the OLA algorithm (maximum~5\%)}
	\label{tbl:olasupp}
        \scriptsize
	\begin{tabular}{|l|rrrrrr|}
\hline
 & \multicolumn{6}{c|}{$k$} \\
{\bf Dataset} & {\bf 2} & {\bf 5} & {\bf 10} & {\bf 20} & {\bf 50} & {\bf 100} \\ \hline
Adult & 1.67\% & 2.42\% & 3.71\% & 1.47\% & 1.76\% & 4.80\%\\
Florida & 4.77\% & 3.91\% & 4.15\% & 4.18\% & 3.47\% & 3.48\%\\
Michigan & 4.83\% & 1.88\% & 3.73\% & 3.96\% & 1.53\% & 1.34\%\\
\hline
	\end{tabular}
\end{table}

\section{Experimental Evaluation} \label{sc:experiments}

The approach presented in this paper needs to be evaluated in terms of risk and information loss.
We measured the impact on information loss based on the metric we described in Section~\ref{sc:eval:iloss}.
For risk, we evaluated the nearest-neighbour linking model (Section~\ref{sc:eval:link}) and measured the suppression rates needed to enforce confidence-based $k$-anonymity (Section~\ref{sc:eval:pseudokanon}).
We also evaluated the computational performance of the model.

We used the {\ke}-anonymisation algorithm on all three datasets: Adult, Florida and Michigan.
All datasets were evaluated using the OLA and Mondrian $k$-anonymisation algorithms and for various $k$ and $\epsilon$ to allow us to examine how the model behaves for different parameters.

\subsection{Impact on Information Loss}

As a first step we tried to understand the relationship between $k$ and $\epsilon$ and the resulting information loss on the $\epsilon$-quasis.
We expect that information loss will increase with decreasing $\epsilon$, inkeeping with (\ref{eq:dberror}).
Additionally, we expect information loss to increase with $k$, due to the increased diameter of the data in larger equivalence classes.

Our results are in line with expectations.
Values of $\epsilon$ of \num{8} and \num{16} consistently produce errors less than 5\%.

\begin{figure}[ht]
	\includegraphics[width=\columnwidth]{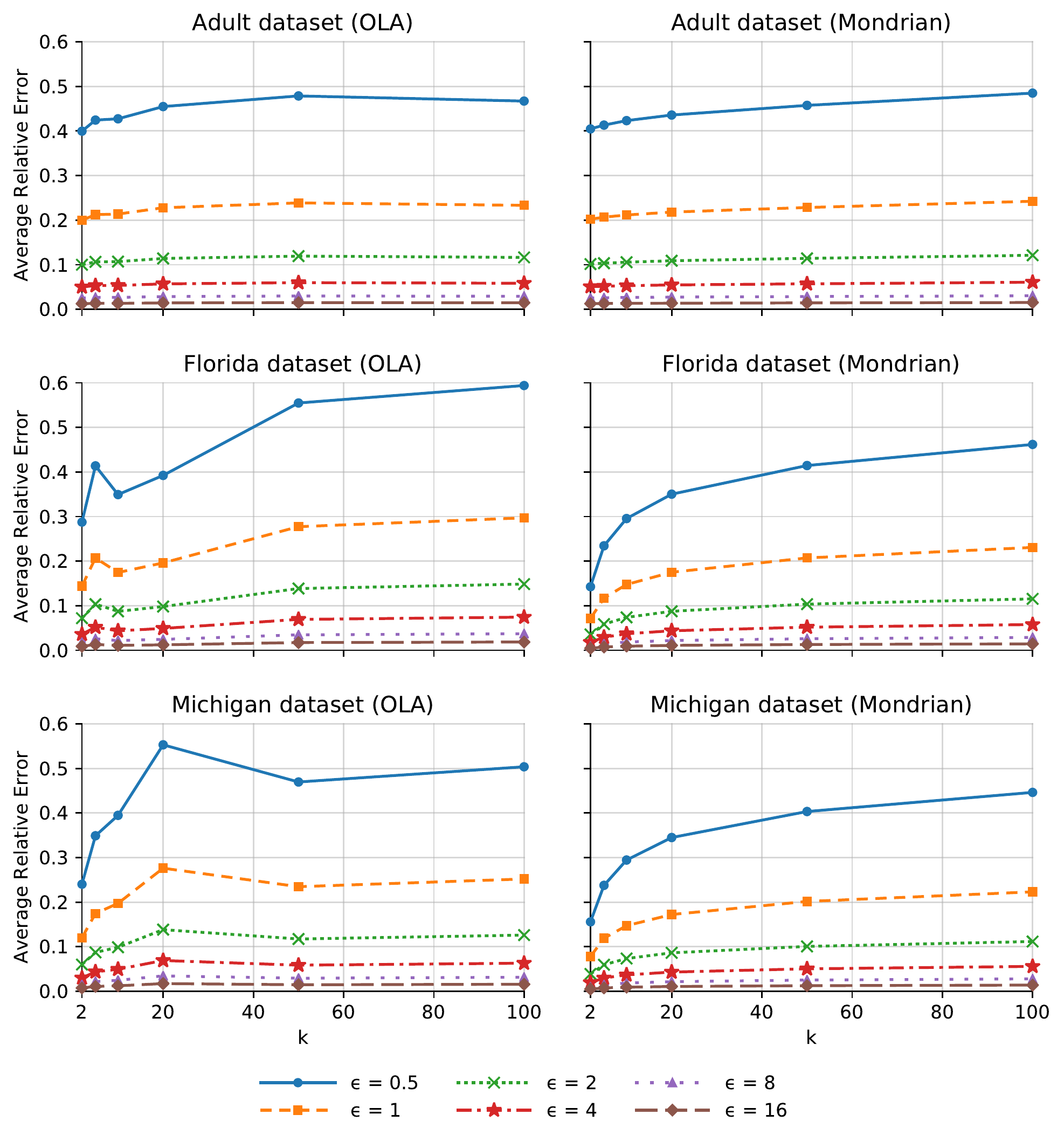}
  \caption{\label{fig:eval:errorvk} Average relative error for the height\slash weight attribute versus $k$ for various datasets, $k$-anonymisation algorithms and $\epsilon$ values.}
\end{figure}

In Figure~\ref{fig:eval:errorvk} the average relative error is plotted against $k$ for various values of $\epsilon$.
As expected, we can see decreasing information loss for our mechanism with decreasing $k$.
This behaviour is most pronounced for the Mondrian algorithm, with small values of $k$ benefitting most from Mondrian's local recoding strategy.

\begin{figure}[ht]
	\includegraphics[width=\columnwidth]{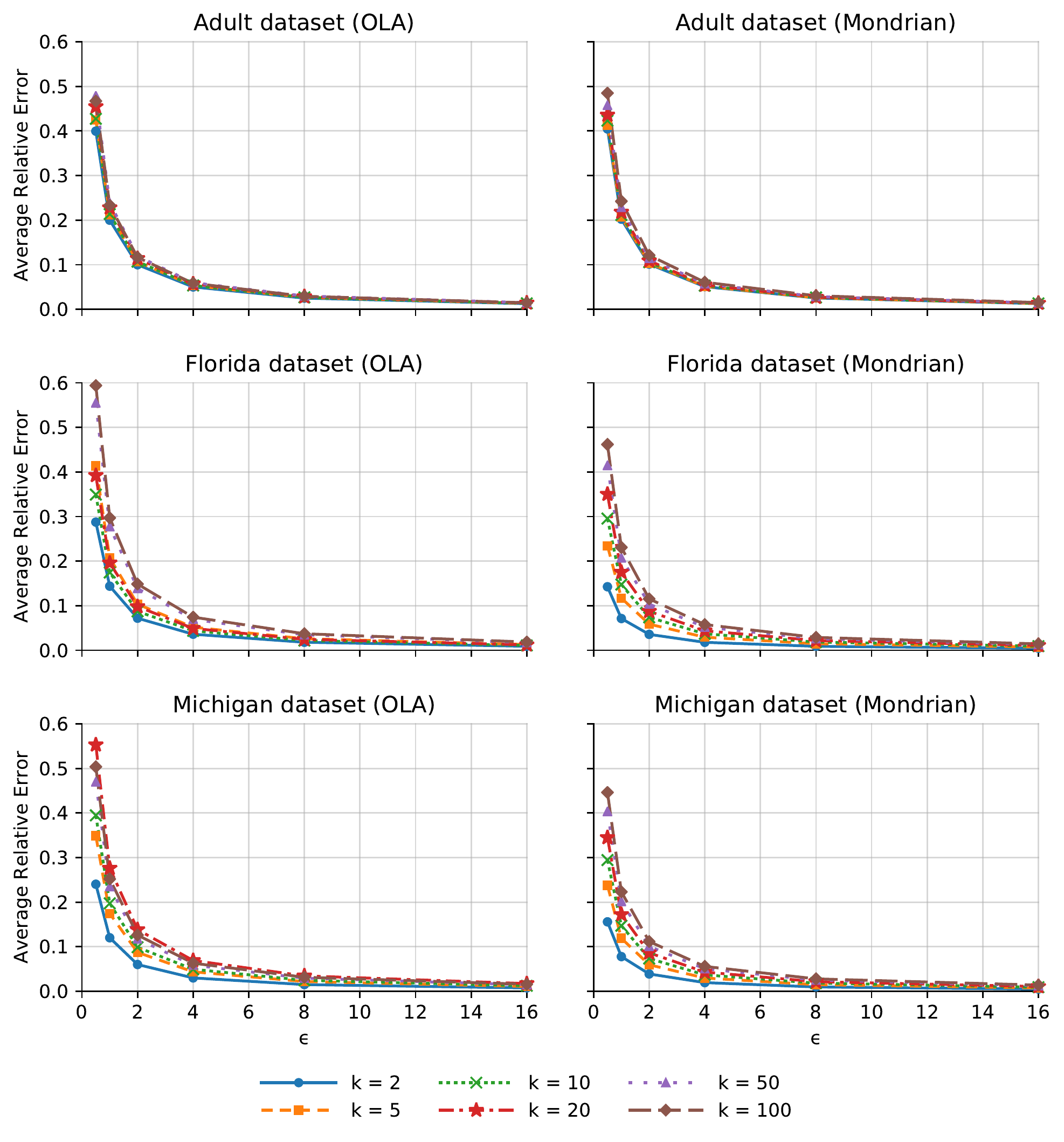}
  \caption{\label{fig:eval:errorveps} Average relative error for the height\slash weight attribute versus $\epsilon$ for various datasets, $k$-anonymisation algorithms and $k$ values.}
\end{figure}

In Figure~\ref{fig:eval:errorveps} the average relative error is plotted against $\epsilon$ for various values of $k$, swapping the plotting order to Figure~\ref{fig:eval:errorvk}.
As expected we see a close correlation between error and $\epsilon$, with errors consistently below 5\% for $\epsilon \ge 8$.
The inversely proportional behaviour corresponds directly to the information loss model (\ref{eq:dberror}), which contains a $\frac{1}{\epsilon}$ term independently of $k$.

We see a number of algorithm-specific discrepancies in Figure~\ref{fig:eval:errorvk} for the OLA algorithm.
For the Adult dataset with OLA, the error is seen to drop for $k=100$.
This can be attributed to the increased suppression imposed by OLA at $k=100$ (4.8\%, see Table~\ref{tbl:olasupp}) compared to $k=50$ (1.76\%).
With greater OLA suppression, more outlying values are suppressed, thereby reducing the diameter of equivalence classes, and hence reducing the noise added by differential privacy.
For the Florida dataset with OLA, the spike at $k=5$ can similarly be explained by suppression, albeit by reduced OLA suppression.
With reduced suppression, fewer outliers are removed, requiring additional noise to be added to the data to anonymise these values.

Figures~\ref{fig:eval:errorvk} and~\ref{fig:eval:errorveps} also demonstrate the greater levels of flexibility afforded to data curators in the tradeoff of privacy (using $\epsilon$) and information loss (using average relative error).

\subsection{Risk Evaluation}

After analysing the impact of the proposed method on information loss, we now try to understand the relationship between selecting $k$ and $\epsilon$ and the resulting linking risk on the dataset.
We expect that the linking risk, as presented in Section~\ref{sc:eval:link}, will decrease with decreasing $\epsilon$ and decrease with increasing $k$.
As $\epsilon$ dictates the strength of privacy (with larger $\epsilon$ guaranteeing less privacy), the linking risk should increase as privacy decreases.
Similarly with $k$, the larger the $k$, the greater the privacy guarantee, hence we expect reduced linking risk when $k$ is increased.

\begin{figure}[ht]
	\includegraphics[width=\columnwidth]{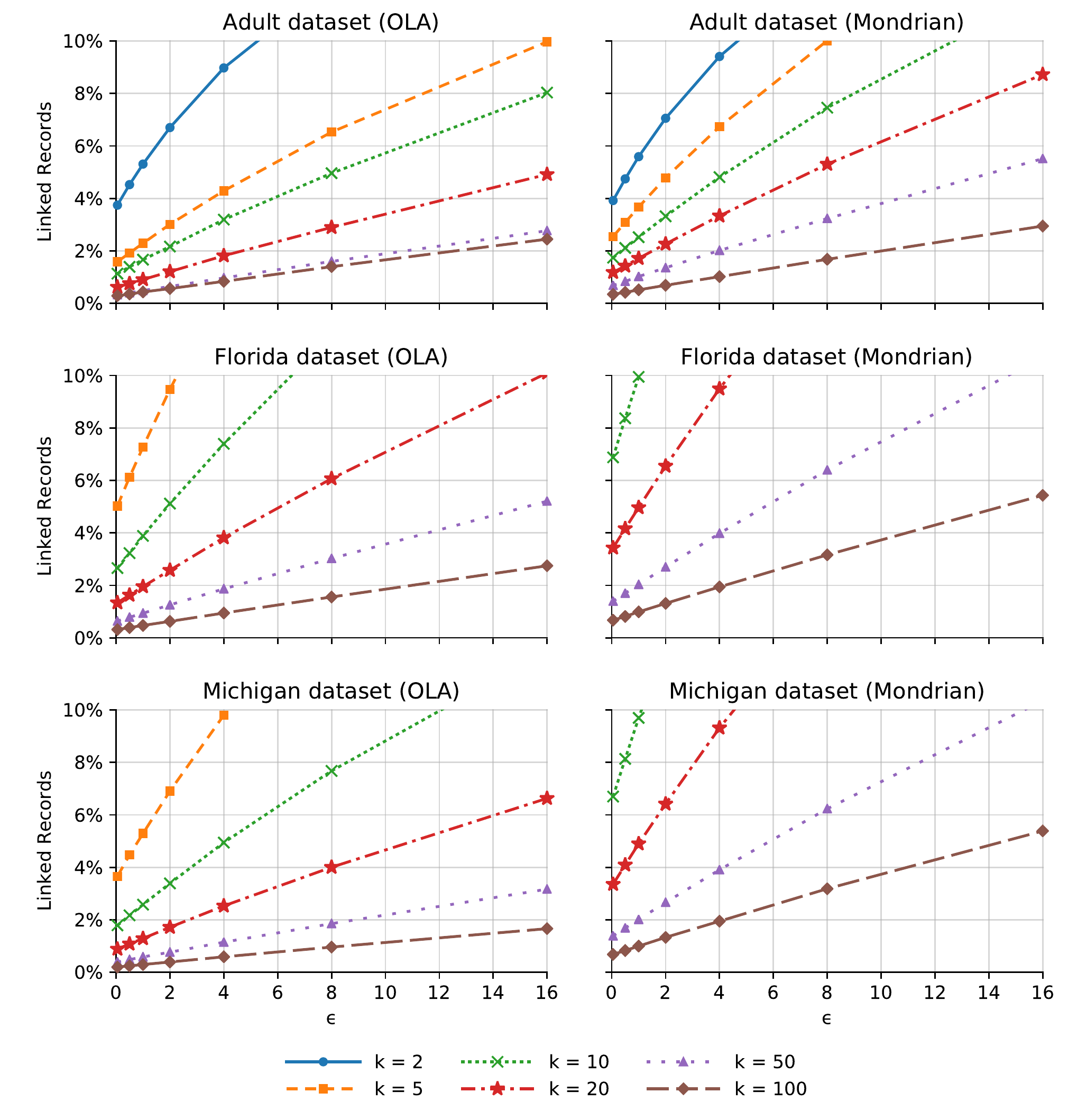}
  \caption{\label{fig:eval:link} Percentage of linked records versus $\epsilon$ for various datasets, $k$-anonymisation algorithms and $k$ values.}
\end{figure}

Figure~\ref{fig:eval:link} shows the linking risk of {\ke}-anonymity, as a percentage of linkable records, versus $\epsilon$ for various values of $k$.
We can see that linking risk below 5\% is attainable for moderate values of $k$ and $\epsilon$ in all the set-ups examined.
Across all datasets, large values of $k$ present linking risk below or close to 5\%, even for larger $\epsilon$.
For smaller values of $k$, the value of $\epsilon$ may need to be tuned, but linking risk below 5\% is still attainable.

We note that the linking risk in some cases exceeds the $k$-anonymisation risk of $\frac{1}{k}$.
This is due to our use of an additional quasi identifier in the linking attack, where this quasi has not be anonymised in line with $k$-anonymity.
The additional information therefore presents additional information to an attacker to help re-identify records.

In line with expectations, linking risk decreases when privacy guarantees are increased.
This behaviour is seen across all three datasets and across both $k$-anonymisation algorithms.
It should be noted that the Florida and Michigan datasets are missing data for small $k$ values, in particular \num{2} and \num{5}.
This is because the linking risk associated with these values of $k$ for these datasets exceeds 10\%, so falls outside the range of the plot.
We decide not to extend the plotting region since we view linking risks over 10\% as excessively risky to be considered for use.

\subsection{Measuring Confidence-Based \texorpdfstring{$k$}{k}-Anonymity}

The last evaluation step is understanding the relationship between $k$ and $\epsilon$ and the suppressions required to achieve confidence-based $k$-anonymity.
By suppressing risky records, the original guarantees of $k$-anonymity can be preserved.
We expect fewer risky records to be found for small $\epsilon$, reflecting the greater privacy guarantees offered by smaller $\epsilon$.

\begin{figure}[tb]
	\includegraphics[width=\columnwidth]{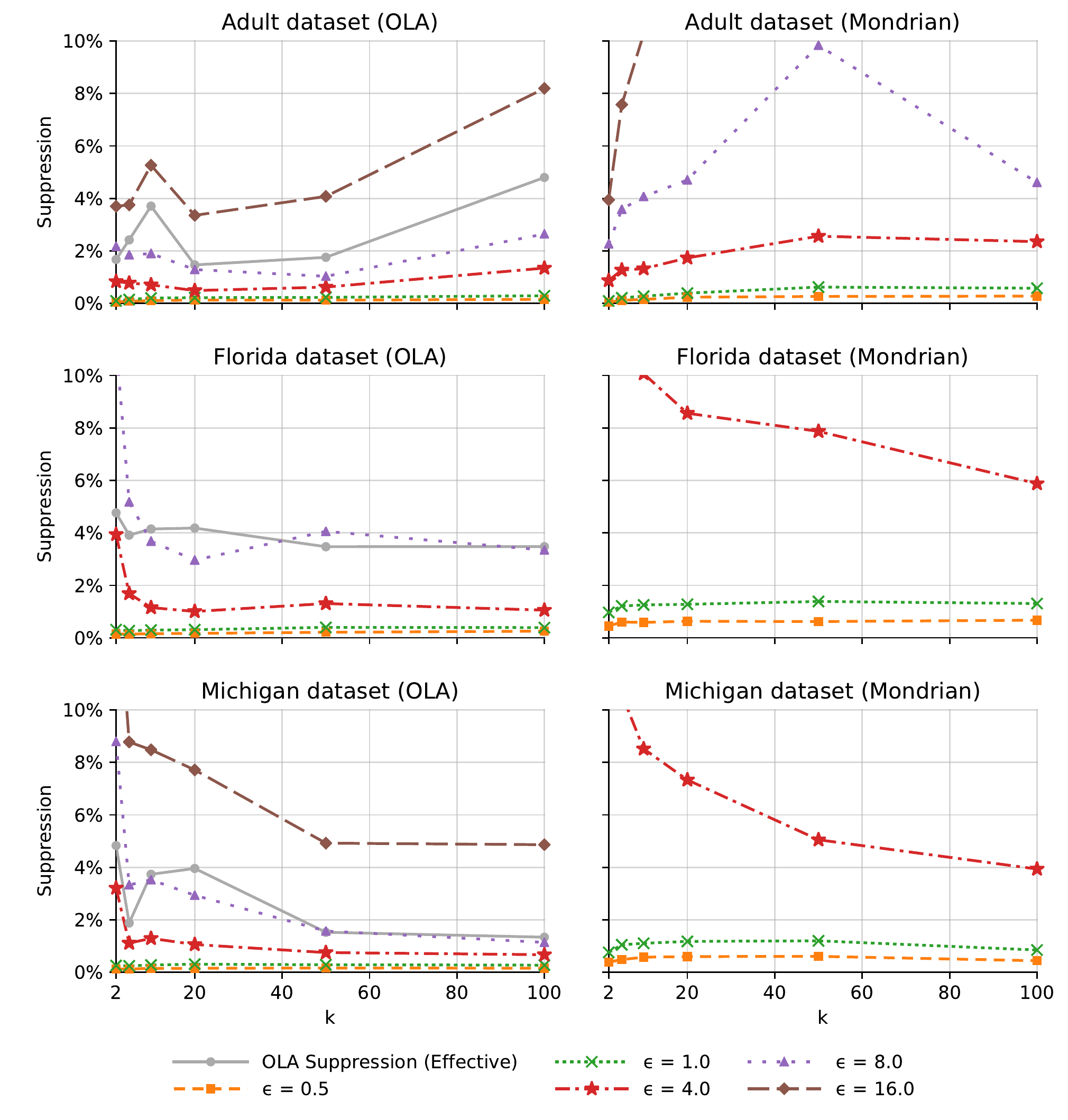}
  \caption{\label{fig:eval:kanon} Percentage of additional suppressions required to maintain \num{0.99}-pseudo $k$-anonymity versus $k$ for various datasets, $k$-anonymisation algorithms and $\epsilon$ values.}
\end{figure}

For the purpose of these evaluations we tested confidence-based $k$-anonymity for confidence values of \num{70}\%, \num{90}\%, \num{99}\%, \num{99.9}\%, \num{99.99}\% and \num{99.999}\%.
In Figure~\ref{fig:eval:kanon} we plot the results for a confidence value of \num{99}\%, which we consider to be a reasonable confidence threshold.
The behaviour for other confidences produces qualitatively similar results, but with different volumes of suppressions.
Greater confidence corresponds to a larger confidence interval, hence imposing fewer suppressions, while more suppressions are required when the confidence is reduced.

For small values of $\epsilon$, 99\%-confident $k$-anonymity can be achieved by suppressing fewer than 2\% of records.
This is evident across all datasets and is independent of $k$.

When using OLA, the algorithm imposes suppressions on the data to achieve $k$-anonymity.
These suppressions are shown in Figure~\ref{fig:eval:kanon} as a solid grey line, and a user-specified maximum of 5\% has been imposed.
Importantly, for small $\epsilon$, 99\%-confident $k$-anonymity is achievable in most cases without exceeding the suppression budget of 5\%, when OLA's suppression and confidence-based $k$-anonymity suppressions are summed.

When using the Mondrian algorithm, the same behaviour is observed for small $\epsilon$, while larger $\epsilon$ values can be used while still keeping suppression below 5\%.

Figure~\ref{fig:eval:kanon} highlights the data-dependence of this model.
The model is particularly sensitive to the distribution of the values in an equivalence class, and to the size of equivalence classes.
Equivalence classes close in size to $k$ are more likely to be suppressed entirely.
Therefore, an dataset and algorithm which produces smaller equivalence classes will result in greater suppression to meet confidence-based $k$-anonymity.
This behaviour can be seen in the increased suppression rates for the Mondrian algorithm against OLA.

Nevertheless, achieving less than 5\% suppression for various $\epsilon$ and $k$ values shows the {\ke}-anonymity model satisfies its goal of protecting the privacy of individuals in the data.

\subsection{System performance}

We evaluated the system performance of {\ke}-anonymity in terms of running time. We compared the execution time of both Mondrian and OLA 
algorithms for the Florida and Michigan datasets when applying $k$-anonymity with $\epsilon$-quasis treated as $k$-quasis against
the execution time of {\ke}-anonymity. 

\begin{figure}[t]
	\subfloat[Florida, Mondrian]{\label{sfloridam}\includegraphics[width=0.22\textwidth]{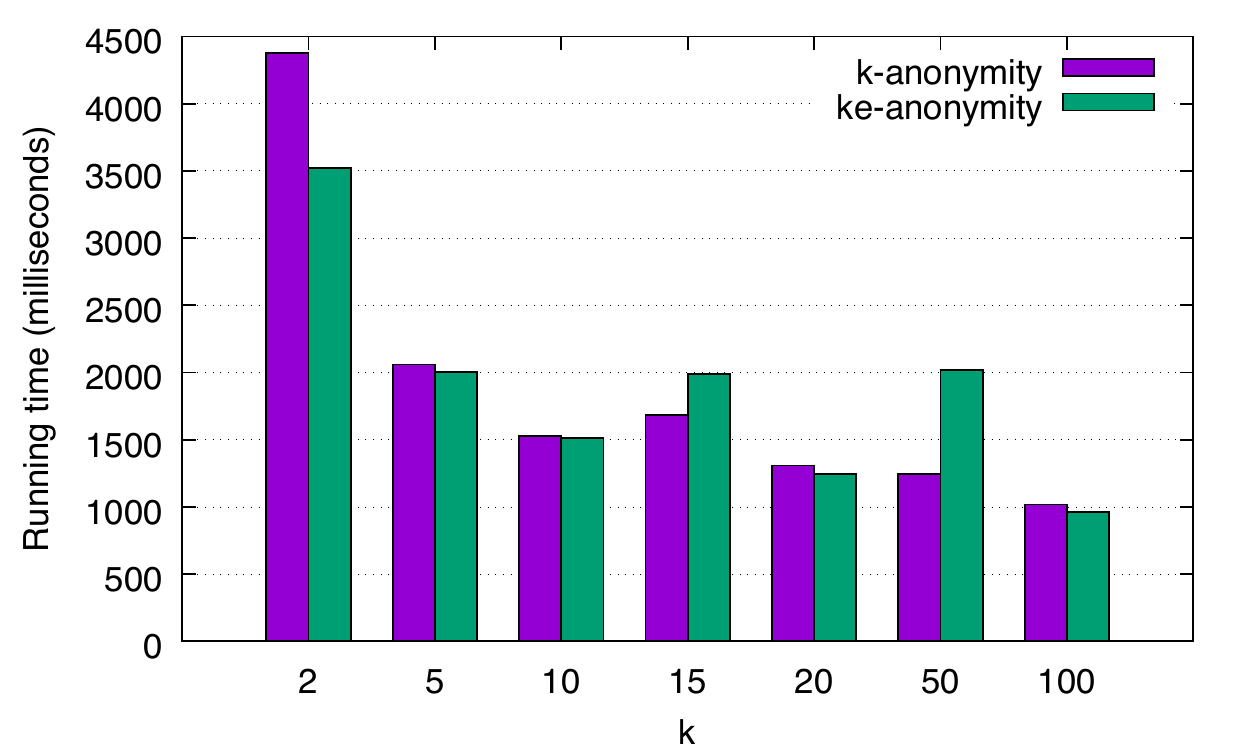}}
  \qquad	
	\subfloat[Michigan, Mondrian]{\label{smichiganm}\includegraphics[width=0.22\textwidth]{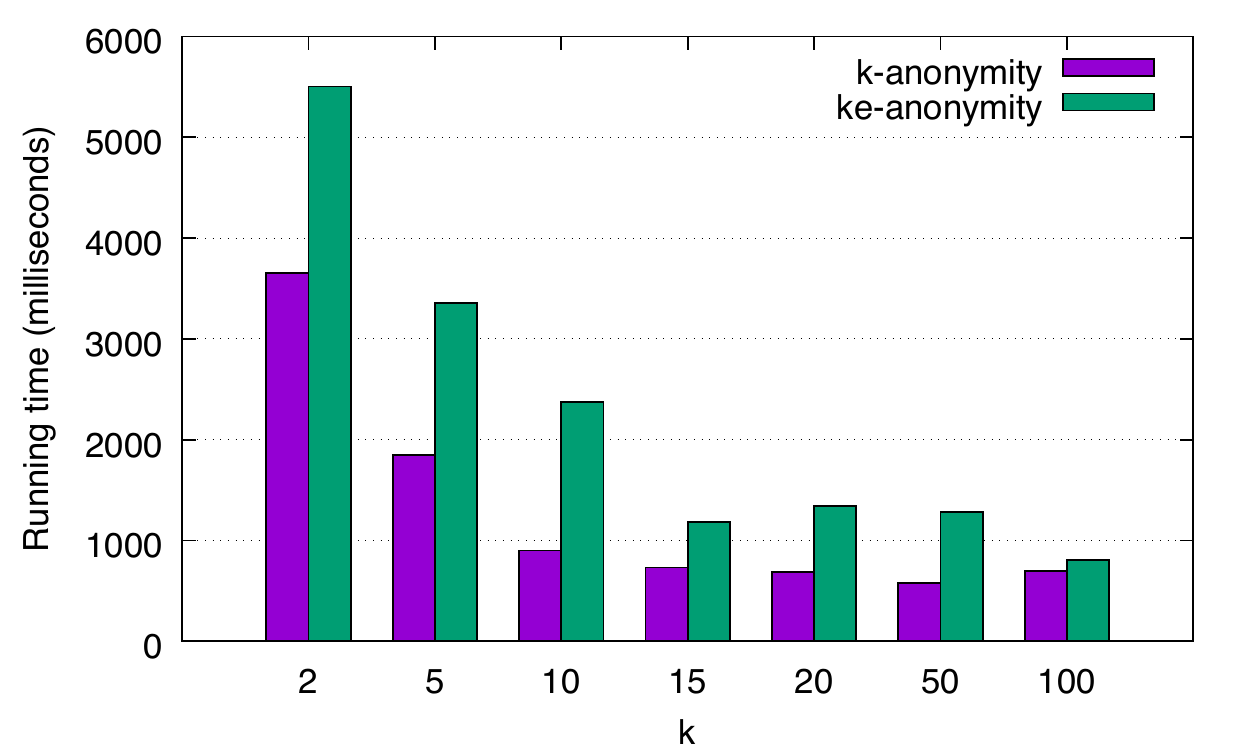}}
  \qquad
	\caption{\label{fig:systemperformancemondrian} Running time of $k$-anonymity versus {\ke}-anonymity when Mondrian is used as the base algorithm}
\end{figure}

\begin{figure}[t]
	\subfloat[Florida, OLA]{\label{sfloridaola}\includegraphics[width=0.22\textwidth]{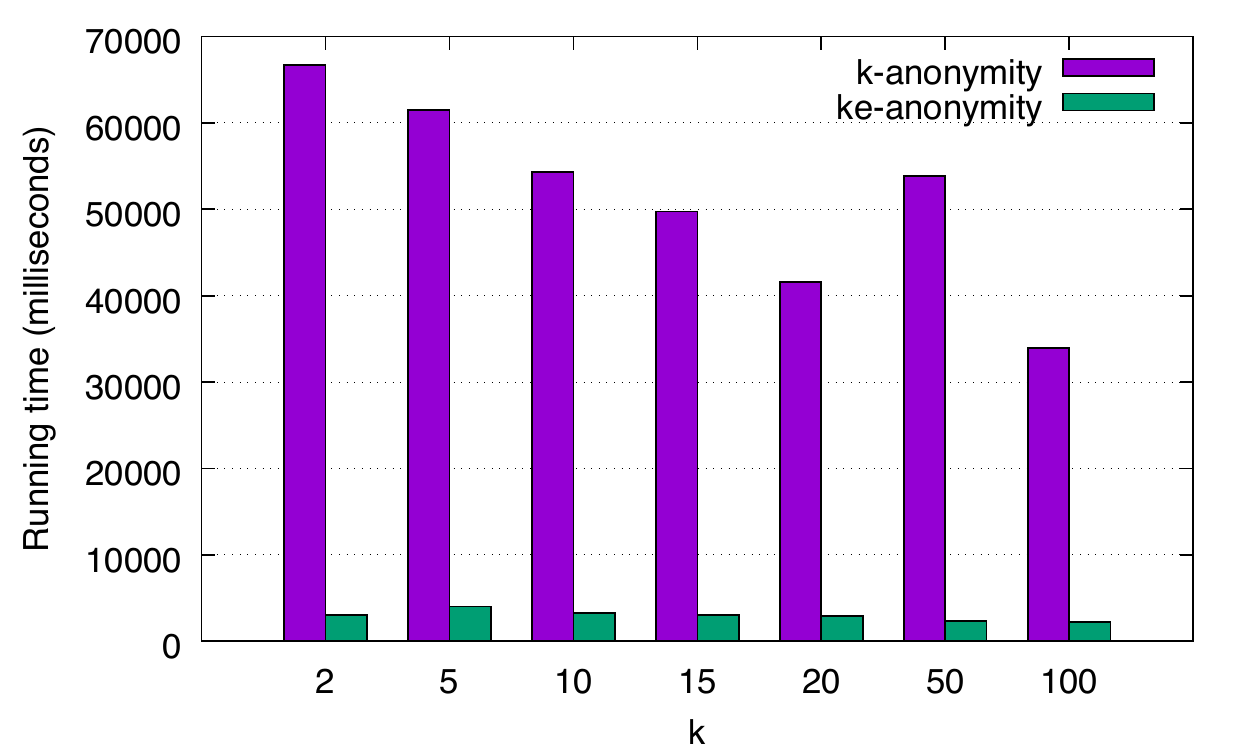}} 
  \qquad	
	\subfloat[Michigan, OLA]{\label{smichiganola}\includegraphics[width=0.22\textwidth]{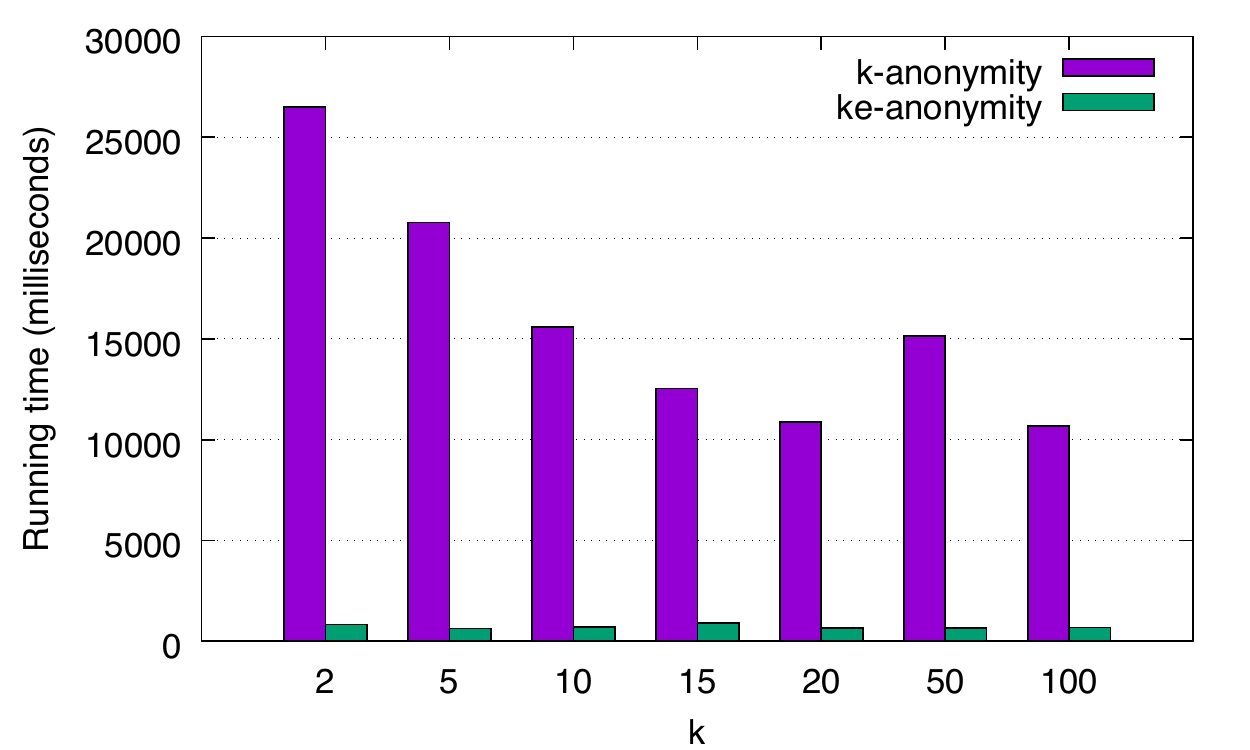}}
	\caption{\label{fig:systemperformanceola} Running time of $k$-anonymity versus {\ke}-anonymity when OLA is used as the base algorithm}
\end{figure}

The results for Mondrian and OLA (with suppression threshold set to 5\%) are shown in Figures \ref{fig:systemperformancemondrian} and \ref{fig:systemperformanceola} respectively. 
We observe that in the case of Mondrian, as displayed at Figure \ref{fig:systemperformancemondrian}, the execution time of 
{\ke}-anonymity increases across all experiments. This is due to the fact that we have to perform an extra loop over the dataset to apply
the $\epsilon$-differential privacy for each equivalence class. On the other hand, for OLA we notice a significant decrease 
in the execution time, ranging from 14x to 32x speedup, as shown in Figure \ref{fig:systemperformanceola}. This performance increase is attributed to the fact that {\ke}-anonymity 
reduces the dimensionality of quasis that need to be explored by OLA. When applying {\ke}-anonymity the exploration lattice for OLA is 8 times smaller and the lattice exploration
converges very fast, absorbing the cost of the extra loop to apply differential privacy. Similar results were observed for the Adult dataset.

\section{Conclusion}\label{sc:conc}

In this paper we presented {\ke}-anonymity, a framework that combines $k$-anonymity with $\epsilon$-differential privacy.
This work lays out the model foundation to combine different anonymisation protocols and present them in a uniform way in terms of information loss and risk. 
Our model was validated experimentally using real datasets and use cases.
The results of our experiments have shown that the combination of the two anonymisation protocols is both feasible and has controllable properties.
We demonstrated the impact on the information loss, where {\ke}-anonymity produces better results than $k$-anonymity alone due to reduced dimensionality of the quasi identifiers space.
We also presented the linking risk and our approach on how to achieve confidence-based $k$-anonymity under a strong adversarial model.
Our results have shown that even for high confidence levels we only need to suppress a small percentage of records in order to achieve the desirable privacy guarantees.
We believe that our work provides a fresh perspective on utilising different anonymisation protocols that, so far, are exclusively used in isolation.

\bibliographystyle{IEEEtran}

\end{document}